\documentclass[aps,reprint,superscriptaddress,
nofootinbib,showpacs,amsmath,amssymb]{revtex4-1}

\usepackage{latexsym,graphicx,slashed,bm,dcolumn}

\newcommand{\bi}{\bibitem}
\newcommand{\be}{\begin{eqnarray}}
\newcommand{\ee}{\end{eqnarray}}
\newcommand{\beqn}{\begin{eqnarray}}
\newcommand{\eeqn}{\end{eqnarray}}
\newcommand{\rar}{\rightarrow}

\newcommand{\mug}{\mu{\rm G}}

\newcommand{\vect}[1]{\mbox{\boldmath$#1$}}

\def\vnabla{\vect{\nabla}}
\def\vtimes{\vect{\times}}
\def\vB{\vect{B}}
\def\vE{\vect{E}}
\def\vF{\vect{F}}

\def\vJ{\vect{J}}
\def\vv{\vect{v}}

\def\vr{\vect{r}}

\begin{document}

\title{Dark matter and generation of galactic magnetic fields}

\author{Zurab~Berezhiani}
\affiliation{Dipartimento di Fisica e Chimica, Universit\`a di L'Aquila, 67100 Coppito, L'Aquila, Italy} 
\affiliation{INFN, Laboratori Nazionali del Gran Sasso, 67010 Assergi,  L'Aquila, Italy}

\author{A.D.~Dolgov}
\affiliation{Novosibirsk State University, Novosibirsk 90, Russia} 
\affiliation{ITEP, Moscow, 113218, Russia} 
\affiliation{Dipartimento di Fisica, Universit\`a di Ferrara, 44124 Ferrara, Italy}
\affiliation{INFN, Sezione di Ferrara, 44124 Ferrara, Italy} 
\author{I.I. Tkachev}
\affiliation{Novosibirsk State University, Novosibirsk 90, Russia} 
\affiliation{Institute of Nuclear Research, 
Moscow, 188300, Russia}

\date{\today}


\begin{abstract}
A  new scenario for creation of galactic  magnetic fields is proposed which is operative at the 
cosmological epoch of the galaxy formation, and  which relies on unconventional properties 
of dark matter. Namely, it requires existence of feeble but long range interaction between the 
dark matter particles and electrons.  In particular, millicharged dark matter particles  or mirror 
particles with the photon kinetic mixing to the usual photon can be considered. We show that 
in rotating protogalaxies circular electric currents  can be generated by the interactions of free 
electrons with dark matter particles in the halo, while the impact of such interactions on galactic 
protons is considerably weaker. The induced currents may be strong enough to create the 
observed magnetic fields on the galaxy scales  with the help of moderate dynamo amplification. 
In addition, the angular momentum transfer from the rotating gas to dark matter component 
could change the dark matter profile and formation of  cusps at galactic centers would be inhibited. 
The global motion of the ionized gas could produce sufficiently large magnetic fields also in 
filaments and galaxy clusters. 
 \end{abstract}

\maketitle

The origin of large scale magnetic fields remains one of very deep cosmological mysteries. 
Magnetic fields are detected in galaxies of all types and they constitute a very important 
component of the galactic dynamics since they are relevant for compression of the gas clouds,  influence star formation process,  and determine the spectrum of  the galactic  cosmic rays. 
The Milky Way, for example, possesses the magnetic  field of a  few $\mug$  over the plane of its disc, with a coherence length of  a few kpc. 
Similar magnetic fields have been detected in other disc galaxies and  
also in high redshift protogalactic structures. 
However, cosmic magnetism is not related only to galaxies: 
observations point to the presence of magnetic fields in galaxy clusters, 
though with weaker strength but on much larger scales. 
Detailed description of the observational data can be found e.g. in reviews~\cite{magn-obs}. 

It is widely accepted that the magnetic fields observed in disk galaxies are enhanced 
by galactic dynamo due to combined effect of differential rotation and helical turbulence. 
However, this paradigm  in itself is incomplete  since {\it per s\`e} 
dynamo mechanism principally cannot explain  the origin of the initial magnetic 
fields acting as seeds. It can only amplify the magnetic field if its initial value was non-zero 
(for reviews, see refs. \cite{dynamo}). No compelling mechanism for formation of large scale seed 
fields has been found yet. Usually the problem is that the strength 
of the seed fields is too weak and their coherence length is too short, so  
huge galactic dynamo is necessary for amplifying it up to observed values.  
The task of creation of intergalactic magnetic fields is much more severe. 
The problem is that the dynamo mechanism is not  efficient on intergalactic scales.

In a wide range of models the seed magnetic fields are generated in the very early universe, at 
different cosmological epochs ranging from inflation to big bang nucleosynthesis 
(for reviews, see e.g. refs.~\cite{magn-theor}). 
The considered mechanisms are based,  generically,  on a new physics outside of the standard cosmological and particle physics frameworks. Some of these mechanisms could provide 
strong enough seeds  to be dynamo-amplified up to the observed galactic fields, 
 but their coherence length is by far smaller than the typical galactic scale. 
Chaotic field line reconnection could stretch up the coherence length 
but at the expense of a strong decrease of the magnitude. 

Magnetic seeds can be generated also at  later cosmological stages, 
from the recombination epoch to the period of  star formation.  
The suggested mechanisms rely on the conventional astrophysics and cosmology 
and might give rise to reasonably strong magnetic seeds  with large coherence scales. 
They employ the Biermann battery effect~\cite{biermann}  
or the turbulence generated at the radiation decoupling era  
and  difference between the drag forces  exerted on electrons and protons  
by the cosmic microwave background (CMB) radiation~\cite{kulsrud,ber-dol,matta-2}. 
Non-zero vorticities induced at the recombination and radiation decoupling 
epoch by the photon diffusion in  second order in the temperature/density  
fluctuations~\cite{ber-dol,matta-2} may generate 
the seeds of future galactic magnetic fields $\sim 10^{-20}$~G 
which in principle could be amplified by galactic dynamo 
up to the observed magnitude $\sim 10^{-6}$~G.

Dynamo mechanism can lead to an exponential increase of the galactic magnetic fields 
till they  reach the saturation  value, $B_{\rm eq}$, of about a few $\mu$G. 
Saturation takes place when equipartition between the magnetic and turbulent energy 
densities is achieved, 
i.e. $\rho_B = B^2/8\pi$ becomes equal to  $\sim\rho v^2$, where
 $\rho$ is the typical matter density of the protogalaxy
 and $v$ is the turbulent velocity in the  system. 
After this moment the further growth of magnetic fields is halted 
by the dynamical back reaction of magnetic stress on the turbulence.
Curiously, the energy density of magnetic field $\rho_B = B^2/8\pi$ 
is equal to the CMB density 
$\rho_{\rm CMB} = 2 \times 10^{-51}$ GeV$^4$ when $B=3~\mug$.  

In order to reach the saturation value during cosmological time $t$, 
the seed fields at the time of the galaxy formation $t_{\rm gal}$ must be big enough, 
$B_{\rm seed} > B_{\rm eq}  \exp[-(t-t_{\rm gal})/ \tau_{\rm dyn}]$, 
where $\tau_{\rm dyn}  \sim 0.2-0.5$ Gyr is a typical e-folding time in the dynamo regime.  
Its precise value strongly depends on the characteristics of the turbulent plasma 
in the protogalaxy as well as on the angular velocity profile, 
but the above estimate seems to be reasonable.  
Hence, for the Universe age $t \simeq 14$~Gyr,  
the galactic dynamo could amplify magnetic fields up to 
$B_{\rm eq}\sim 10^{-6}$~G starting from 
$B_{\rm seed}\sim 10^{-20}$~G or even less. 
Nevertheless, the situation cannot be considered satisfactory.  
There is a mounting evidence that magnetic fields in and around normal galaxies were  
already of the order of $\mug$ in galaxies at large cosmological redshifts  $z\sim 1\div 2$, 
which are too young for an efficient  galactic dynamo, since
at this time the Universe was only about one-third of its present age~\cite{Kronberg:2007dy}.  
This observation significantly reduces the number of available e-foldings 
and thus requires much stronger magnetic seeds. 
E.g. for a galaxy formed at $z=6$ or $t_{\rm gal} = 1$~Gyr,  
in order to reach  $B_{\rm eq} \sim 10^{-6}$~G at the cosmological time 
$t \approx 4.4$~Gyr (corresponding to redshift  $z =1.5$),  
the seed fields at least of the order of $10^{-15}$~G are required 
even  if the e-folding time is taken as $\tau_{\rm dyn} = 0.2$~Gyr.

In addition, all disk galaxies, whenever appropriate observations are available, 
show that the regular (mean) component of the magnetic fields is nearly the same 
as the random (RMS) component which indicates that  the coherence length of the seeds 
should be comparable to the galaxy scale~\cite{dynamo}.  
Summarizing these observations, it seems difficult to avoid the conclusion 
that the magnetic seeds at epoch of galaxy formation must be rather large, 
$B_{\rm seed} > 10^{-15}$~G,  with the coherence scale at least of the order of 1 kpc. 

Note also that the radio observations of magnetic fields in the edge-on spiral galaxies 
suggest that mostly the dominant component of the magnetic field is parallel to the disk plane \cite{Dumke}. 
However, for some galaxies magnetic fields have strong vertical components extending 
far away from the disk plane \cite{Hummel}, which may indicate that  the dynamo had worked 
for a relatively short time and large poloidal component of the magnetic field was maintained. 

\medskip 

In this paper we propose a new mechanism which leads to generation of 
rather strong magnetic seeds during the process of the galaxy formation. This 
mechanism has interesting implications for the nature of dark matter. 
In a sense this mechanism is a generalization of our previous work~\cite{ber-dol} 
to the epoch when galaxies or protogalaxies were already created and the 
vorticity perturbations evolved up to significantly  high values. 
At this stage circular electric currents  could be generated  in rotating protogalaxies 
due to different drag forces  exerted on protons and electrons by the CMB radiation. 
However, the  seed magnetic fields generated in this way are uncomfortably low 
even for galaxies, to say nothing of the galactic clusters. 
The situation can be significantly improved if there exist relatively light dark matter particles 
which have some feeble but long range interaction with electrons.  
In this case, the friction force, produced  by the dark matter particle interactions with
the electron-proton plasma in the galactic halo, which is directed opposite to the galaxy rotation, 
causes a drag of electrons relative to ions,  while the latter, 
along with the neutral atoms and molecules, rotate as a whole galaxy. 
Hence a circular electric current would be induced.
This is an essence of our  proposal. In the most optimistic case even 
the intergalactic magnetic fields of the proper strength can be generated 
with a moderate dynamo application. 
Thus, if such particles exist, the long standing problem of generation 
of galactic and intergalactic magnetic fields can be settled down.

\medskip 
As we know, galaxies start to form at cosmological redshift $ z \sim 10$ or so, 
which corresponds to the cosmological time 
$t_z \simeq (2/3) H_0^{-1} [\Omega_m (1+z)^3]^{-1/2} \sim 0.5$~Gyr, 
though most of the large galaxies are formed at much lower redshifts. 
The high density regions 
where primordial fluctuations have grown sufficiently large, 
start to collapse once their self-gravity begins to dominate over the cosmological expansion. 
The  thermal bremsstrahlung cooling and dissipative contraction, recombination,   
fragmentation in molecular hydrogen clouds,  and finally star production  lead to creation of galaxies. 
Tidal torques acting between density fluctuations lead to galaxy  rotation. 
On the other hand, observations show that the universe was reionized  
around the same epoch, in the redshift interval $z \sim 10\div 6$, 
presumably by the ultraviolet emission from the first generation of massive stars 
and/or quasars.  

For a simple order of magnitude estimate let us consider a protogalaxy as a cloud 
of partially ionized matter which rotates in the isotropic sea of the CMB photons  
with temperature $T = (1+z) T_0$, where $T_0= 2.73$~K.  
Let us denote  the fraction of the ionized matter as $\xi$, so  
the number density of free electrons (ions) is $n_e = \xi n_B$, 
where $n_B$ 
is the number density of baryons in the cloud. 
For  simplicity, we assume that all ions are protons, neglecting contribution of helium. 
Apart from the regular rotational  velocity $v_{\rm rot}$, 
electrons and protons have chaotic thermal velocities, 
$v_{e,p} \sim  (T_{e,p}/m_{e,p})^{1/2}$. Usually one takes $T_p=T_e$.

The CMB photons interact with electrons with the Thomson cross-section 
$\sigma_{e\gamma} = 8\pi\alpha^2/3m_e^2 = 6.65 \times 10^{-25}~{\rm cm}^2$, 
where $m_e$ is the electron mass and $\alpha = e^2=1/137$ (we use the CGS system of units). 
Thus, the regular part of the  drag force induced by the CMB on the electron 
in plasma with a local flow velocity $\vv$ is 
\begin{equation} \label{F-e}
\vF = e \vv  B_F
\end{equation}
where factor $B_F$ can be estimated as 
\beqn \label{B-gamma} 
B_F =  \sigma_{e\gamma}^{~} n_\gamma \omega_\gamma/e  
= 3.4 \times 10^{-30}\, (1+z)^4 ~{\rm eV}^2/e
\nonumber \\ 
= 5.8 \times 10^{-28}\, (1+z)^4 ~{\rm G} \, .
\eeqn
Here  $n_\gamma$ and $\omega_\gamma$ 
 are respectively  the number density of the CMB photons  and  their mean energy 
 at the cosmological epoch corresponding to redshift $z$. 
The CMB pressure on protons is completely negligible since the respective 
cross-section is  smaller than $\sigma_{e\gamma}$  by the factor $(m_p/m_e)^2$. 
Therefore,  in the rotating protogalaxy the friction force created by the CMB pressure 
causes a relative drag of electrons in the direction opposite to the galaxy rotation,
while the ions rotate  as a whole galaxy. Hence, circular currents must be  
induced, remarkably without charge displacement and local violation of the electric neutrality. 

An important parameter governing the magnitude of this current is  the electric conductivity, 
which in turn is determined by the Coulomb scattering among  the electrons and ions (protons). 
Namely, the CMB pressure induces coherent (de)acceleration of electrons with respect to protons 
but the coherence is destroyed by their chaotic collisions due to their thermal motion.  
The characteristic collision time, $\tau_{ep}$, due to the Coulomb scattering 
between $e$ and $p$  can be estimated as
\begin{equation} 
\tau_{ep} = \frac{m_e^2 \langle v_e^2 \rangle }{4\pi  \alpha^2 \langle 1/v_e \rangle n_e L_e} 
\simeq  \frac{m_e^{1/2} T_e^{3/2}}{ 4\pi \alpha^2 n_e L_{e}} \, ,
\label{nu}
\end{equation}
where $L_e  \sim 10$ is the electron Coulomb logarithm.  
In the above equation, we have taken the line-of-sight 
thermal average as $\langle v_e^2 \rangle = T_e/m_e$. 
Therefore, for the electric conductivity of the plasma  we have 
\begin{equation}
\sigma =  \frac{e^2 n_e  \tau_{ep}}{2 m_e}   
\simeq    \frac{T_e^{3/2}/m_e^{1/2} }{  8\pi \alpha  L_e }  
\sim  10^{12}~{\rm s}^{-1}  
\left( \frac{T_e}{10^4 \,{\rm K}} \right)^{3/2}  
\label{sigma}
\end{equation}
Note, that it does not depend on the density of charge carriers, $n_e$, i.e. on the 
ionization degree $\xi$, unless the latter is so small that the resistance is dominated 
by neutral atoms.  

Due to the radiation drag force (\ref{F-e})  electrons slow down with respect to ions, while 
the latter  keeps on rotating with velocity $v_{\rm rot}$ practically non-attenuated. 
For the difference between the mean rotational velocities of 
ions and electrons we find $\Delta v_e = \tau_{ep} F /2m_e \ll v_{\rm rot}$ 
which induces circular electric currents with density   
$j = e n_e \Delta v_e=\sigma F/e=\sigma v_{\rm rot} B_F$. 
One could naively estimate the magnetic field generated by the CMB induced current  in a rotating galaxy via the Biot-Savart law as $B \sim 4\pi j R = {\cal R}_M B_F$, 
where $R$ is the galaxy radius and ${\cal R}_M \equiv 4 \pi\sigma v_{\rm rot} R$, i.e.   
\begin{equation} \label{RM} 
{\cal R}_M 
\simeq  \frac{4\times 10^{22}}{L_e }    
\left(\frac{R \, v_{\rm rot} }{ 10^3~{\rm kpc \cdot km/s}}\right)  
\left( \frac{T_e}{10^4~ {\rm K}} \right)^{3/2} . 
\end{equation}
For a typical galaxy with $R\sim 10$ kpc and rotational velocity 
$v_{\rm rot} \sim 100$ km/s  this would result in quite large magnetic field.   
In view of eqs. (\ref{B-gamma}) and  (\ref{RM}), the magnitude of 
$B\sim {\cal R}_M B_F$  can reach 1 $\mu$G,  
which is practically the observed value of magnetic fields in galaxies.

However, the Biot-Savart law is valid only when the stationary regime is reached, 
while the system under scrutiny is far from that. 
The time to reach the stationary situation is much longer than the cosmological time. 
To see that let us consider the Maxwell equations in the cosmological plasma 
and modification of the MHD equations in presence of extra non-potential forces 
related to a dark matter interaction with electrons. 
Namely,  let us consider the electric current $\vJ = \sigma (\vE +  \vv \times \vB + \vF/e)$,  
where $\vF$ is the external  force acting  on electrons, see Eq.~(\ref{F-e}).  
In our case it is the drag force induced by the interaction with the CMB 
(or with dark matter halo, see below). 
Finding electric  field $\vE$ from this equation and 
substituting it into equation  $\partial_t \vB = - \vnabla \times \vE$, 
we obtain $\partial_t \vB = \vnabla \times (\vv \times \vB + \vF/e - \vJ / \sigma)$.  
Substituting the above expression for $\vJ$  and using 
$\vnabla \times \vE =- \partial_t \vB$ and $\vnabla \cdot \vB =0$, we come to
\begin{equation} \label{MHD}
 \partial_t \vB =  \vnabla\! \times \! \vF/e \, + \, \vnabla \! \times \! (\vv \! \times \! \vB) \, + \, 
 \frac{1}{4\pi\sigma} (\Delta \vB +  \partial^2 _t \vB),  
\end{equation} 
which is in fact the MHD equation  in  the presence of external source
term $\vnabla \! \times \! \vF/e= B_F \vnabla \! \times \vv \, + \, (\vnabla B_F)\times \vv$. 
In the limit of high conductivity, the second term in the MHD equation, 
the so called advection term, leads to a dynamo effect on the magnetic seed fields 
once the value of the latter is non-zero. 
It is well-known, however, that in absence of the source term, 
the MHD equations cannot give rise to non-zero magnetic field if $\vB=0$ initially.

In our case, assuming $\vB=0$ at $t=0$, we find that the source term (\ref{B-gamma}) 
induces a nonzero magnetic seed field which initially grows 
roughly as
\be\label{B-fin}
\vB(t) =  \int_{0}^t  dt \, \vnabla \! \times \! \vF/e  
=  \int_{0}^t dt \,   \vnabla \! \times \! (B_F \vv) \,.
\ee
However, taking into account  that $B_F(t) \propto (1+z)^4 \propto t^{-8/3}$, see eq. (\ref{B-gamma}), 
we find that the biggest value for the magnetic seeds can be obtained around the 
cosmological epoch of hydrogen recombination and photon decoupling, 
$z  \sim 1000$, or $t  \sim 5 \times 10^5$~yr. 
At earlier times, the plasma is strongly coupled and the relative motion of the electrons 
with respect to protons is negligible, hence the lower limit of the integration is irrelevant.  
The mean value of vorticity at the characteristic spatial scale $\lambda$ is  
$\Omega_\lambda = | \vnabla \! \times \! \vv |_\lambda  \leq  10^3(\delta T/T)^2/\lambda$, 
as calculated in Ref. \cite{ber-dol}. Then one can estimate of a seed field
generated at this epoch, say with  the coherence length  $\lambda$ of few kpc 
which corresponds to the present day comoving scale of a typical galaxy $\sim 1$ Mpc, 
as $B_\lambda \sim \Omega_\lambda t_{\rm rec} B_F(t_{\rm rec}) \leq 10^{-20}$~G,  
in agreement with the results of refs.~\cite{ber-dol,matta-2}.
However, as it was emphasized  earlier, such magnitude of the seeds is still too small.
The seed fields  with the coherence length $\sim 1$ kpc and  
$B_{\rm seed} >  10^{-15}$~G are needed to conform to observations of  
coherent magnetic fields $\sim \mug$ at high redshift galaxies \cite{Kronberg:2007dy}.

\medskip  

In what follows we consider the formation of galactic magnetic fields  
by some hypothetical dark matter particles, $X$, 
forming extended halos around the galaxies. 
We note that drag force exerted on electrons by $X$ particles is given again 
by eq. (\ref{F-e}) but with $B_F$ proportional to 
their number density  in the halo, to their average momentum, and 
of  to the cross-section of their elastic scattering on electrons. 
Therefore, to produce sufficient pressure on electrons, larger than that of CMB,  
the latter cross-section must be large enough at low momentum transfer. 
This is feasible if these $X$ particles have long range interaction with normal matter. 
As a natural example, we consider millicharged particles 
in the mass range from several keV  to several GeV. 

Millicharged particles, that can be  either bosons or fermions, 
have tiny electric charges  $e' = \epsilon e $, 
where $e$ is the electron charge and $\epsilon \ll 1$,  which are bounded 
by direct laboratory limits as well as by cosmological and astrophysical  observations. 
These bounds strongly depend on the masses of $X$-particles.   
If $X$-particles are lighter than electron, $m_X < m_e$, a strong laboratory limit 
comes from the bounds on the invisible decay of ortho-positronium into $X\bar X$, 
according to which $e' < 3.4\cdot 10^{-5} e$ \cite{positronium}. 
For $m_X < 1$ keV, a stronger bound, $e' < 10^{-5} e$ was obtained in ref. 
\cite{Gninenko:2006fi} but we shall not consider so small masses here. 
For $m_X > m_e$ the direct bounds were obtained in ref.~\cite{Prinz}. 
For $m_X = 1$ MeV these bounds give $e'/e < 4.1 \times 10^{-5}$ while for larger masses 
they become weaker, as e.g. $e'/e < 5.8 \times 10^{-4}$ for $m_X=100$ MeV. 
For $X$-particles heavier than 100 MeV, even  $e' \sim 10^{-2} e$ is allowed 
whereas for $m_X > 1$ GeV, $e'$ can be as large as $e/10$.  
More bounds on the mass/coupling of millicharged particles, 
derived from astrophysics and cosmology 
can be found in refs.~\cite{raffelt,milli-limits}.
In the following we assume that  $m_X > 10$ keV, in order 
to avoid strong limits on $e'$  from stellar evolution. 

The presence of light millicharged particles, with $m_X < 1$ MeV or so,  
during the epoch of big bang nucleosynthesis (BBN) influences standard 
cosmological picture in several respects. 
In particular, the expansion rate of the Universe and 
baryon-to-photon ratio can be significantly 
altered during BBN and this may be dangerous \cite{milli-limits}. 
The impact of such particles on BBN is discussed in our paper~\cite{BDT-BBN}, 
with the conclusion that the model is not ruled out especially if the lepton asymmetry 
is non-zero. 

For a simple consistent model for the millicharged $X$ particles,  
one may consider a hidden gauge sector of particles,  
which, among other possible gauge factors,  contains $U(1)$ gauge group
 and respective `photon' field $A'_\mu$ interacting with $X$ particles.  
This paraphoton $A'_\mu$  may have a kinetic mixing, 
$({\epsilon}/{2}) F'_{\mu\nu} F^{\mu\nu}$,   to ordinary photon $A_\mu$ \cite{Holdom}. 
The prototype model is given e.g. by asymmetric mirror world, a shadow sector of the particles  
having strong and electroweak interactions  similar to the ordinary particles, 
 but with the electroweak and QCD scales $v'_W$ and $\Lambda'$ being different 
 from the ordinary ones \cite{BDM}. 
Lightest stable particles of this sector are
$e'$ and $p'$ with opposite electric-like charges of a shadow 
$U(1)'$ gauge group. They resemble  our electron and proton  
and their stability is guaranteed by conservation of respective `baryon' number. 
Thus, such  parallel matter would be dark matter for us. 
Kinetic mixing of shadow and ordinary $U(1)$ gauge factors 
makes shadow electron $e'$ and proton $p'$ to be millicharged 
(with electric charges $\sim \epsilon$) with respect to our photon 
and gives rise to effective long range interactions between the ordinary and dark particles.\footnote{
Mirror world with microphysics exactly identical to 
the Standard Model also provides a viable dark matter (see a review \cite{Alice} and references 
therein). However, in this case the positronium oscillation limit 
on the photon--mirror photon kinetic mixing is very strong, 
$\epsilon < 4 \times 10^{-7}$ \cite{positronium}
and there  are cosmological limits two orders of magnitude more restrictive \cite{Lepidi}. 
In the case of asymmetric mirror sector \cite{BDM}, with $m'_e > m_e$, the positronium limits are
irrelevant, and the cosmological limits are also more flexible \cite{Lepidi}. }
Kinetic mixing parameter $\epsilon$ can be considered as a (field-dependent) 
dynamical degree of freedom, with interesting implications for the time variability 
of the fine structure constant $\alpha$ \cite{Savely}.

For additional simplification, we can assume that the direct 
product $U(1) \times U(1)'$ of two gauge factors is spontaneously broken down to 
some diagonal $U(1)_{\rm em}$ representing the true massless photon eigenstate $\gamma$ 
while another mass eigenstate $\gamma'$ becomes heavy. This true photon may predominantly interact
with ordinary particles with the coupling constant $e$, but it would see shadow 
particles as having electric charges $e'= \epsilon e$. Clearly, such a theory has 
no anomalies. Another possibility is to  leave the shadow photon massless as well,  
directly in spirit of ref. \cite{Holdom}.

These two cases have different cosmological implications. 
Namely, if $X$ particles emerge from a shadow world with  massless $\gamma'$, 
then $X\bar X \to \gamma' \gamma'$ annihilation must be essential and 
the frozen cosmological number density of $X$-particles (i.e. mirror `electrons' $e'$) 
most probably would be determined by their charge asymmetry \cite{BDM,Alice} 
as is a natural case of asymmetric dark matter. 
Such particles could recombine with their counterparts of the opposite sign of their mirror 
electric charge (mirror `protons' $p'$), so  after  their recombination the density 
of charged mirror particles may be rather small. 
Rich spectrum of other possibilities will be considered elsewhere.

\medskip 

In the subsequent estimates we assume that  $\gamma'$ is heavy
and hence $X$-particles have no relevant self-interactions, 
except for only feeble interactions  with the normal matter induced by their millicharges 
with respect to our photon $\gamma$. In this case  they practically  do 
not recombine because their binding energy is tiny and the Bohr radius is huge.

The processes like $e^+e^- \to X \bar X$ induced 
via the photon kinetic mixing bring $X$ particles into equilibrium 
with ordinary matter below temperatures $T\sim \epsilon^2 \alpha^2 M_{Pl}$.  
Then, once  they were thermally produced in the early universe, 
their  present abundance is given by the expression~\cite{frozen}:
\be
\Omega_X h^2   \approx 0.023 ~ x_f \, g_{\ast f}^{-1/2} 
\left(\frac {v\sigma_{\rm ann}}{ 1~{\rm pb} }\right )^{-1}, 
\label{n-X}
\ee
where $ \sigma_{\rm ann} $ is the $X\bar X$ annihilation cross section,  and 
$x_f \equiv m_X/T_f = 10 + \ln(g_X/g_{\ast f}) + \ln m_{\rm MeV} +0.5 \ln x_f$ 
is the $X$-particle mass ratio to the annihilation freezing temperature, with 
$g_X$ being the number of 
the degrees of freedom of $X$-particle, $ g_{\ast f}$ being the effective number of degrees 
of freedom of all particles in the plasma at $T_f$, and $m_{\rm MeV} = m_X/{\rm MeV}$.  Typically $x_f \sim 10$;  
it varies between $5\div 20$ for $X$-particle masses in keV - GeV range.

\medskip 

Let us  first consider the case when $X$-particles are lighter than electrons, $m_X < m_e$. 
Then they can annihilate only into  photons,  with the cross-section 
\be
v \sigma (X \bar X \rar 2\gamma) = \frac{\pi \alpha^{\prime 2} }{ m_X^2} = 
\frac{\epsilon_5^4}{m_{\rm keV}^2} \times 6.5 \cdot 10^{-4} ~ {\rm pb},
\label{sigma-X-antiX-gamma}
\ee
where $m_{\rm keV} = m_X/{\rm keV}$,  
$\alpha' = e'^2 /4\pi = \epsilon^2 \alpha$ and $\epsilon_5 = 10^{5} \epsilon$. 
Therefore, for $X$ particles in the mass range of several keV, their 
cosmological energy density would be
\be
\Omega_X h^2   \approx 150 \times \left( \frac {m_{\rm keV}}{\epsilon_5^2 } \right)^2 .
\label{rho-X-light}
\ee
Therefore, for $m_X > 10$ keV the dark matter 
density would be  overproduced unless $\epsilon_5 > 20$ or so, which contradicts to 
the positronium bound $\epsilon_5  < 3.4$ \cite{positronium}.\footnote{
One can assume, however, that $\epsilon$ is changing with time. In particular, 
its  initial value can be larger  than $2\times 10^{-4}$ in early epochs, 
effective  for annihilation of $X$-particles, 
while it drops below the positronium limit  today \cite{Savely}. } 
However, one can envisage an additional annihilation channel   
into some lighter species of dark sector (e.g. into mirror neutrinos $\nu'\bar \nu'$ \cite{BDM}) 
which could diminish  $\Omega_X h^2$ down to acceptable values. On the other hand, 
bounds imposed by the CMB anisotropies
do not allow light $X$-particles  to constitute the dominant part of dark matter, 
and they may provide only its rather small fraction,  
with a conservative bound 
$\Omega_X h^2 < 0.007$    \cite{Dubovsky:2003yn}. 
Hence, the presence of some other form of dark matter is also needed.
In what follows, we take the fraction $\Omega_X$
with respect to the total mass density of dark matter as a free parameter.

The origin of the above bound is the following. 
The collision time of $X$-particles with respect to  $eX$-scattering in the primeval plasma 
 is given by eq.~(\ref{nu}), where one has to substitute $m_X< m_e$ instead 
 of $m_e$, $v_X> v_e $ instead of $v_e$, and $\alpha \alpha'$ instead of $\alpha^2$:
\be
\tau_{eX} = \frac{m_X^2 v_X^3 }{ 4\pi \alpha \alpha' n_e\, L} ~ . 
\label{l-free-cosm}
\ee 
The number density of free electrons is  
$n_e = \xi n_B = 2.5 \times 10^{-7} \xi(z) (1+z)^3~{\rm cm}^{-3}$, 
where $\xi(z)$ is the ionization degree at redshift $z$ and 
$\eta = n_B/n_\gamma = 6\times 10^{-10}$ is the baryon to photon ratio.  
Just before the hydrogen recombination, $T >0.2$~eV or $z > 1100$,  when $\xi = 1$, 
the chaotic velocity of $X$-particles is $v_X \sim (T/m_X)^{1/2}$,  
and the $Xe$-collision time,  
$\tau_{eX} \simeq 1.6  \times 10^9 \, (m_{\rm keV}^{1/2}/\epsilon_5^2)(1100/z)^{3/2}$~s,  
is shorter than the cosmological time, $t = 2H^{-1}/3 = 1.6 \times 10^{13} (1100/z)^{3/2}$~s, 
unless parameter $\epsilon$ is very small 
($\tau_{eX}/t \simeq (10^{-4}/\xi) (m_{\rm keV}^{1/2}/\epsilon_5^2) < 1$).   
 
So, prior to recombination $X$-particles are strongly coupled to electrons and through them to 
photons, and thus cannot participate in the structure formation.  
After recombination, when $\xi$ drops down to $10^{-4}$, the collision time rises up 
and becomes bigger than the cosmological time, 
so $X$-particles decouple from the usual matter and do not follow the bulk  motion of the 
baryon matter. 
At this stage $X$-particles behave as a warm dark matter component 
and participate in the large scale structure formation along with the rest of dark matter. 
The bound $\Omega_X h^2 < 0.007$ was 
obtained in ref. \cite{Dubovsky:2003yn} on the basis of the early WMAP results. 
Today, in light of high precision data acquired by the Planck Collaboration, 
this bound should become more stringent and deserves an independent study.

Let us return to the question of magnetic field generation in the process of galaxy formation
and consider  a partially ionized baryon cloud of typical galactic mass 
$M \sim 10^{11}~M_\odot$. At e.g. $z \sim 6$ such an object, 
undergoing contraction due to thermal cooling, 
would have radial size $R\sim 100$ kpc, 
the baryon overdensity $\kappa_B \sim 10^2$, 
and  temperature $T_e\sim 10^3$ K, hence $v_e = (T_e/m_e)^{1/2} \sim 10^{-3}c$. 
As for $X$-particles,  
they would be distributed in halo as a dark matter component, 
with typical virial velocity $v_X$.

The drag force exerted by $X$-particles on electrons is given again by eq. (\ref{F-e}), 
but with factor $B_F$ proportional to the number density of $X$-particles, 
to their momentum, and to their elastic scattering cross-section on electrons:   
\be
B_F = \sigma_{eX} n_X  m_X v_{\rm rel}/e 
\label{BF-X}
\ee
with $v_{\rm rel}$ being the relative velocity 
between the electrons and $X$ particles. 
Namely, $v_{\rm rel}\simeq v_X$, if $v_X > v_e$. 
In the opposite case one has to substitute $v_X$ by $v_e$. 
We have 
\be
v_{\rm rel} \sigma_{eX} = \frac{4\pi \alpha \alpha' L}{m_X^2 v_{\rm rel}^3} 
\simeq \frac{  2 \cdot 10^{-8}\, \epsilon_5^2 }{m_{\rm keV}^{2} v_{100}^3 } 
 ~ {\rm eV}^{-2} ,
\label{sigma-X}
\ee
where $L\sim 10$ is the corresponding Coulomb logarithm and 
$v_{100} = (v_{\rm rel}/100~{\rm km/s})$.  
The mass density of $X$-particles is equal to: 
\be
m_X n_X \approx  \Omega_X h^2  (1+z)^3 \kappa  \times 10^{-10}~{\rm eV}^4 ,
\label{nXmX}
\ee
where $\kappa (z)$ is the dark matter overdensity in the galactic halo with respect 
to its mean cosmological density at redshift $z$. 
Thus, we obtain
\be
B_F \simeq \frac{\epsilon_5^2 }{m_{\rm keV}^2 }\, 
\frac{\Omega_X h^2 \,  \kappa  (1+z)^3}{v_{100}^3}  \times 4 \cdot 10^{-16}\,{\rm G} 
\label{B-X}
\ee
(c.f. eq. (\ref{B-gamma})). 
Let us note again that $B_F$ practically does not depend on the 
ionization degree  of matter  
unless it becomes so small that the resistance is dominated by neutral particles. 
So, this estimate is valid for  the post-recombination residual ionization degree $\xi \sim 10^{-4}$.

Now, before evaluating the strength of the magnetic field generated by dark matter in the galactic haloes, 
let us discuss the following. In the case of the current generated 
by the CMB photons we have taken into consideration 
 that the (proto)galactic matter rotates, while the CMB photons do not. 
 It should be verified if the same assumption is valid for $X$-particles. 
 The collision time of $X$-particles with electrons and protons,  
  given by eq.~(\ref{l-free-cosm}), at the galaxy formation epoch reads  
\be
\tau_{eX} \approx \frac{m_{\rm keV}^2 v_{100}^3 }{\epsilon_5^2 \kappa_B \xi  (1+z)^3 } \times 
2 \cdot 10^{14}~ {\rm s}
\label{tau-of-T-m}
\ee 
Therefore, taking $m_X \sim10$ keV and $\epsilon \sim 10^{-5}$, we 
see that for the galaxy forming at redshift $z\simeq 6$ (i.e. $t \simeq 1$ Gyr), 
with $\kappa_B\sim 10^2$ for the baryon overdensity, we get $\tau_{eX}\geq t$,  
if $\xi \leq 10^{-2}$.
Therefore, at this epoch the bulk motion of $X$ component could be 
 again independent from that of the normal matter. 
For  later epochs, with further increasing $\kappa_B$ and $\xi$,  
we have $\tau_{eX}< t$, so the $X$ particles would be dragged by the rotation 
of the normal matter.\footnote{Presumably, gradual reionization of the Universe started 
at $z\simeq 10$ whereas at $z< 6$ it was already completely ionized. 
Notice that for  $m_X \sim100$ keV, we get $\tau_{eX}\geq 1$ Gyr 
even if $\xi =1$. 
} 


Now we can estimate the magnitude of the
magnetic field generated via the $X$-particle pressure on the 
electrons in the process of galaxy formation. 
For the collapsing protogalaxy, where $B_F$ is not decreasing in time,   
the integral (\ref{B-fin}) is saturated 
on the upper limit, therefore the integration time can be formally as 
large as the  age of the Universe. 
However, the Universe was completely reionized at $z\sim 6$  and this would diminish 
relative velocities among normal matter and $X$ particles because of 
their mutual interactions (c.f. (\ref{tau-of-T-m})). 
Thus, to avoid the discussion of the corresponding subtleties let us integrate up to redshift 
$z=6$, when the cosmological age was about 1~Gyr.  
Therefore, integrating eq.~(\ref{B-fin}) with $B_F $ given by (\ref{B-X}) up to 
$t = 1$ Gyr, taking $R \sim 100$ kpc,  $\kappa \sim 10^2$, $v_{100} \sim 1$, and
the rotational velocity 
$v_{\rm rot} \sim 10$ km/s and taking into account 
the limit $\Omega_X h^2 < 0.007$, one can estimate  
the value of the galactic magnetic seed at $t=1$~Gyr as
\be
B \sim B_F v_{\rm rot} \frac{t}{R}  
\sim \frac{\epsilon_5^2 } {m_{\rm keV}^2} \times 10^{-14} \,{\rm G}.
\label{pobeda}
\ee
Thus, taking $m_X = 10$ keV and $\epsilon_5 = 3$, 
 one can achieve $B\sim 10^{-15}$~G.
 In addition, 
accounting for the adiabatic rise of magnetic field by a factor of $100$ 
when the protogalaxy shrinks from 
the original 100 kpc down to contemporary galactic size 10 kpc, 
can finally get a galactic magnetic seed $\sim 10^{-13}$~G  without great difficulties.
This estimate exceeds the minimal necessary 
value for the galactic magnetic seeds by several orders of 
magnitude.~\footnote{ Let us remark that in our scenario the definition of the seed field 
is somewhat ambiguous as far as the dynamo mechanism, 
related to the advection term  in eq. (\ref{MHD}) is also at work. 
 Magnetic Reynolds number ${\cal R}_M = 4\pi \sigma v \lambda $ is much larger than 1. 
 Therefore, once the source term  generates non-zero magnetic fields oriented towards local voriticities, this term becomes relevant whereas 
 the diffusion term (the last term in Eq. (\ref{MHD})) can be ignored. 
 Notice that $\vnabla \times (\vv \vtimes \vB)$ becomes comparable to 
 the source term $\vnabla \times (B_F \vv)$ when $B \sim B_F$. } 

\medskip

The situation is quite different when $X$ particles are heavier than electrons. 
If $m_X >m_e$, the dominant annihilation channel becomes  
$X\bar X \rar e^+e^-$, with cross-section  
\be
v \sigma (X \bar X \rar e^+e^-) = \frac{\pi \alpha \alpha' }{ m_X^2} =
\frac{\epsilon_5^2}{m_{\rm MeV}^2} \times 6.5~ {\rm pb} .
\label{sigma-X-antiX-ee}
\ee
Hence, assuming no coannihilations, 
cosmological abundance of $X$-particles reads 
\be
\Omega_X  h^2 = 0.012 \times \left( \frac{m_{\rm MeV}}{\epsilon_{5}} \right)^2  
\label{rho-X-heavy}
\ee
Namely, one can obtain $\Omega_Xh^2 \simeq 0.12$ 
e.g. taking $m_X \simeq 10$ MeV and $\epsilon\simeq  3\times 10^{-5}$, 
or $m_X \simeq 1$ GeV and $\epsilon \simeq 3\times 10^{-3}$, 
in agreement with the experimental bounds of ref. \cite{Prinz}.\footnote{One could  take  
 $m_X \simeq 10$ GeV and $\epsilon \simeq  0.03$, also in agreement 
 with the bounds of ref. \cite{Prinz}. However, that heavy  $X$  particles with so large couplings  
 could be excluded, or already excluded,  by the existing experiments on direct search of dark matter.} 
In the case when $m_X \gg m_e$, 
dark matter of the universe could consist entirely of $X$-particles since  
the CMB bounds of ref. \cite{Dubovsky:2003yn} are not applicable. 
Nevertheless, we still keep $\Omega_Xh^2$ as a free parameter, 
assuming that there might exist other components of dark matter. 
In view of eq. (\ref{rho-X-heavy}), to avoid overproducing of dark matter, 
we must require  that $\epsilon_5/m_{\rm MeV} > 0.3$ or so. 
However, if  there are some additional annihlation channels, this constraint can be also 
circumvented.  In any case, in the process of the galaxy formation the 
density distribution of $X$ particles should have the typical profile of halo 
formed by dark matter.

Now, once $m_X > m_e$ one must substitute 
$m_e$ instead of $m_X$ in eq. (\ref{sigma-X}): 
\be
v_{\rm rel} \sigma_{eX} = \frac{4\pi \alpha \alpha' L}{m_e^2 v_{\rm rel}^3} 
\simeq \frac{\epsilon_5^2 }{v_{100}^3 } 
\times 10^{-13} ~ {\rm eV}^{-2} 
\label{sigma-X-heavy}
\ee
and the drag factor  $B_F$ (\ref{BF-X}) now becomes
\be
B_F \simeq 
\frac{\epsilon_5^2 }{m_{\rm MeV} }\, \frac{\Omega_X h^2 \, \kappa (1+z)^3 }{v_{100}^3}\,
\times 10^{-21}\,{\rm G} 
\label{B-X-heavy}
\ee
Taking into account eq. (\ref{rho-X-heavy}), this can be rewritten as\footnote{
Let us remind, however, that eq. (\ref{rho-X-heavy}) is not valid if 
$X$ particles have additional annihilation channels.}
\be
B_F \simeq  m_{\rm MeV} \, \frac{ \kappa (1+z)^3 }{v_{100}^3}\,
\times 10^{-23}\,{\rm G} 
\label{B-X-heavy-2}
\ee
Now,  $X$-particles, being heavy enough, can saturate the total 
dark matter density of the Universe, i.e. $\Omega_X h^2 \approx 0.12$ 
is allowed, 
which according to (\ref{rho-X-heavy}) yields $(\epsilon_5/m_{\rm MeV})^2 \simeq 0.1$. 
In addition, in this case, electron motion would not be involved  into the galaxy 
rotation after reionization, and thus the effective 
integration time in eq. (\ref{B-fin}) can be taken larger than 1 Gyr  
when the Universe was completely reionized.  
Namely, integrating up to $t \simeq 2$ Gyr, assuming that for this 
time the shape of galaxy was already practically settled 
to its present form, with say $R\simeq 10$ kpc, 
$v_{\rm rot} \simeq 100$ km/s and overdensity $\kappa \sim 10^5$, 
one can achieve for the galactic seed fields 
\be
B \sim B_F v_{\rm rot} \frac{t}{R}  
\sim m_{\rm MeV} \times 10^{-15} \,{\rm G}.
\label{pobeda2}
\ee
Therefore, for $m_X \simeq 1$~MeV, one can 
obtain the galactic seed field $\sim 10^{-15}$~G in a natural way, 
whereas for $m_X \simeq 1$~GeV, the seed magnitude can be as large as $10^{-12}$~G.  
Larger masses require larger values of $\epsilon$, with interesting implications for the 
experimental search of millicharged particles as well as for low threshold experiments 
on direct search of dark matter.

\medskip

The following features are also worth mentioning. 
During evolution of the protogalaxy 
the scattering of electrons and protons on $X$-particles can become strong. 
In particular,  after reionization of the Universe ($\xi = 1$)
the collision time of $X$ particles with protons 
\beqn
\tau_{pX} \approx 
\frac{m_{\rm MeV}^2 v_{100}^3 }{\epsilon_5^2 \kappa_B \xi  (1+z)^3 } \times 
2 \cdot 10^{20}~ {\rm s}  \nonumber \\ 
\approx \frac{ \Omega_X h^2}{0.12} \, \frac{v_{100}^3 }{\kappa_B (1+z)^3 } \times 
2 \cdot 10^{21}~ {\rm s} 
\label{tau-pX}
\eeqn
becomes less than the cosmological time as soon as the effective baryon 
overdensity $\kappa_B$ becomes large enough.\footnote{
We use here eq. (\ref{rho-X-heavy}). 
Then  $\tau_{pX}$ does not depend on $m_X$ and $\epsilon$, but only on 
the cosmological density of $X$ particles, $\Omega_X h^2$.
}  
This would lead to a partial transfer of angular momentum from the rotating ordinary matter to 
dark component in the inner dense regions of galaxies,  avoiding thus the formation 
of the cusp and providing more shallow inner profiles of dark matter halos. 
However, in the external part of the halos, outside the galactic disk where the baryon 
density is small, the collision time must be large and hence the halos should maintain 
their normal density distribution. 
In addition, co-rotation of dark matter in the Galaxy at the position of the sun 
may have interesting implications for direct experimental search of dark matter. 

 \medskip 

Another interesting possibility is related  to the
annihilation of $X$-particles with the mass in the range of a few MeV 
in the galactic center, $X \bar X \to e^+ e^-$,   
which can be the origin of the 511 keV line observed by INTEGRAL/SPI \cite{Integral}. 
Indeed, the rate of $ e^+ e^-$ production in the Galaxy is 
$\int   \langle v_{\rm rel} \sigma (X \bar X \to e^+ e^-)(\vr) \rangle n_X(\vr) dV $, 
where $n_X(\vr)$ the number density of dark $X$ particles at the position $\vr$ 
in the Galaxy.  According to dedicated studies \cite{Boehm:2003bt}, the observed 
flux of 511 keV photons can be explained by dark matter annihilation in 
a dark matter halo with a mild enough inner profile provided that 
\be
\left(\frac{\Omega_X h^2}{m_{\rm MeV}} \right)^2 \! 
\frac{ v\sigma(X \bar X \to e^+ e^-) }{1~{\rm pb} } 
 \simeq (0.5\div 1.5) \cdot 10^{-5}
\label{Boehm} 
\ee
Taking into account eqs. (\ref{sigma-X-antiX-ee}) and (\ref{rho-X-heavy}), 
we find that dependence on $m_X$ disappear from condition~(\ref{Boehm})
and thus it is reduced to  $1/\epsilon_5^2 =  0.005-0.016$, 
or $\epsilon = (0.8\div 1.4) \times 10^{-4}$. 
These values for  the millicharge of $X$ particle are compatible 
with the experimental limits~\cite{Prinz} respectively for $m_X > 5 \div 10$ MeV.  
On the other hand, for e.g. $\epsilon = 8 \times 10^{-5}$, $X$ particles 
could constitute the dark matter density, $\Omega_X h^2 \simeq 0.12$, 
if $m_X \sim 25$~MeV or so (c.f. (\ref{rho-X-heavy})). 
For lighter $X$ particles we have  $\Omega_X h^2 < 0.12$ 
and thus the presence of some other type of dark matter would be required.  

\medskip

Until now we did not discuss whether our mechanism could generate also 
the intergalactic magnetic fields, namely the ones observed in galaxy clusters.  
As we know, the normal matter in clusters is presented dominantly in the form of 
hot gas, and the fraction of the luminous matter (galaxies and stars) is smaller.  
The global motion of the ionized gas  relative to dark $X$ matter and 
induced drag of electrons could produce magnetic fields also in filaments 
and galaxy clusters. 
 This process can be somewhat less efficient 
 than in galaxies, due to smaller overdensity of dark matter  
  and weaker adiabatic amplification. 
  Applying naively the same estimate as (\ref{pobeda2}) for magnetic seeds in clusters 
  but with $R\sim 1$ Mpc, one can achieve magnetic seeds $\sim 10^{-14}$~G.  
  One has to keep in mind, however, but the whole picture can be more complicated, 
 e.g  smaller scale peculiar turbulent motion of the intergalactic gas 
 should be taken into account which could result in larger values of 
 $B$ at the cluster scales than those based on the simplified estimates 
 presented above. This question requires a special investigation. 

\medskip

Concluding, we suggested  a new scenario for generation 
of galactic  magnetic fields which is based on dark matter interaction with the normal 
matter in the process of the galaxy formation. The necessary interaction  
can be  induced by pressure of millicharged dark matter particles on electrons. 
In this case,  circular electric currents  can be generated 
due to rotation of free electrons together with the bulk of normal matter in the galaxy,  
colliding with dark matter particles virialized in the halo.  
The impact of such collisions on galactic protons is considerably weaker due to their
larger mass.  The induced currents may be strong enough to create the observed magnetic fields 
on the galaxy scales  with the help of very moderate dynamo amplification.  
In particular, this can naturally explain the recent observational data which 
suggest that the environments of galaxies were significantly 
magnetized at high redshifts, with magnetic fields that were at least 
as strong already after a few Gyr of the cosmological time, as they are today.

The millicharged dark matter particles  may also have interesting implications for the 
properties of galaxies. Namely, in the  dense inner regions of galaxies, 
the angular momentum transfer from the rotating gas to dark matter component 
could involve the latter into its rotation which would  change the dark matter inner profile 
and prevent the formation of  cusps at galactic centers.  
In addition, 
the observed intensity and shape of 511 keV shining from the galactic core can be 
explained by $X \bar X \to e^+ e^-$ annihilation   
provided that $X$ particles have masses in the range of 5-10 MeV, 
and respectively millicharges in the range  $(0.8\div 1.3) \times 10^{-4}$. 

We conclude that it is quite possible that the mechanism propoesed in this work may 
simultaneously explain both galactic and intergalactic magnetic fields 
by a single hypothesis of existence of millicharged dark matter particles. 
The suggested mechanism has interesting implications for direct experimental search of dark matter 
as well as for a laboratory search of millicharged particles.  
Cosmological consequences of existence of
mirror or mirror-like dark sectors \cite{BDM,Alice}, 
which possess dark massless para-photon and can form a sort of atomic dark matter, 
will be discussed elsewhere.


\vspace{6mm} 
\noindent {\bf Acknowledgements} 
\vspace{2mm}

Z.B. thanks V. Berezhiani for useful discussions. 
A.D. and  I.T. acknowledge  support of the Russian Federation Government Grant 
No. 11.G34.31.0047. 
Work of Z.B. was supported t in part by 
Rustaveli National Science Foundation grant No. DI/8/6-100/12
and in part by the grant N14.U02.21.0913 
of the RF Ministry of Science and Education. 
Z.B. acknowledges hospitality of the Galileo Galilei Institute for Theoretical Physics  
during  the Workshop "Beyond the Standard Model after the first run of the LHC" 
where this work was completed.


\end{document}